\documentclass[aps,prd,
nofootinbib,
preprintnumbers,
twocolumn,
showpacs, 
floatfix,
superscriptaddress
]{revtex4}

\pdfoutput=1

\usepackage{color}
\usepackage[pdftex]{graphicx}

\renewcommand{\div}{\mbox{\rm div }}
\newcommand{\grad}{\mbox{\rm grad }}
\newcommand{\const}{\mbox{\rm const}}

\begin{document}

\title{On tidal capture of primordial black holes by neutron stars}

 \author{Guillaume Defillon}
\affiliation{Service de Physique Th\'{e}orique, 
Universit\'{e} Libre de Bruxelles\\CP225 Boulevard du 
Triomphe, B-1050 Bruxelles, Belgium}
\affiliation{D\'epartement de Physique, Ecole Polytechnique\\
 91128 Palaiseau Cedex, France}

\author{Etienne Granet}
\affiliation{Service de Physique Th\'{e}orique, 
Universit\'{e} Libre de Bruxelles\\CP225 Boulevard du 
Triomphe, B-1050 Bruxelles, Belgium}
\affiliation{D\'epartement de Physique, Ecole Polytechnique\\
 91128 Palaiseau Cedex, France}

\author{Peter Tinyakov}
\affiliation{Service de Physique Th\'{e}orique, 
Universit\'{e} Libre de Bruxelles\\CP225 Boulevard du 
Triomphe, B-1050 Bruxelles, Belgium}

\author{Michel H.G. Tytgat}
\affiliation{Service de Physique Th\'{e}orique,
Universit\'{e} Libre de Bruxelles\\CP225 Boulevard du 
Triomphe, B-1050 Bruxelles, Belgium}

\begin{abstract}

The fraction of primordial black holes (PBHs) of masses $10^{17} - 10^{26}$~g
in the total amount of dark matter may be constrained by considering their
capture by neutron stars (NSs), which leads to the rapid destruction of the
latter. The constraints depend crucially on the capture rate which, in turn,
is determined by the energy loss by a PBH passing through a NS. Two
alternative approaches to estimate the energy loss have been used in the
literature: the one based on the dynamical friction mechanism, and another on
tidal deformations of the NS by the PBH. The second mechanism was claimed to
be more efficient by several orders of magnitude due to the excitation of
particular oscillation modes reminiscent of the surface waves. We address this
disagreement by considering a simple analytically solvable model that consists
of a flat incompressible fluid in an external gravitational field. In this
model, we calculate the energy loss by a PBH traversing the fluid surface. We
find that the excitation of modes with the propagation velocity smaller than
that of PBH is suppressed, which implies that in a realistic situation of a
supersonic PBH the large contributions from the surface waves are absent and
the above two approaches lead to consistent expressions for the energy loss.
\end{abstract}

\maketitle

\section{Introduction} 

One of the prominent candidates to the role of the dark matter (DM) of the
Universe is primordial black holes (PBH)
\cite{Hawking:1971ei,Carr:1974nx,Carr:1975qj}.  PBH could have been created in
the early Universe by a variety of mechanisms
\cite{Frampton:2010sw,Hawkins:2011qz,Green:2014faa} and could have survived
till present epoch comprising all or a fraction of the cold DM. An attractive
feature of this scenario is that it does not require the existence of a new
stable fundamental particle.

PBH are characterized by a single parameter --- the mass. Depending on the
production mechanism, this parameter can take essentially any value ranging
from the Planck mass to many solar masses. In some range of masses the
fraction of PBH in the total amount of DM is constrained by
observations. Barring the possibility of stable remnants of a Planckian mass,
the PBH lighter than $10^{14}$~g cannot survive till present time due to
Hawking evaporation \cite{Hawking:1974rv}. The PBH of slightly higher masses,
up to $\sim 10^{16}$~g, evaporate too efficiently and overproduce the
$\gamma$-ray background radiation \cite{Page:1976wx}. In the mass range
$10^{26}-10^{34}$~g the fraction of PBH is constrained by lensing
\cite{Tisserand:2006zx,Alcock:1998fx}, while at even larger masses the DM
completely consisting of PBH is excluded by the microwave observations
\cite{Ricotti:2007au}.

In the intermediate mass range $10^{17} - 10^{26}$~g the PBH are of a
microscopic size (from nuclear size to a fraction of a millimeter), and are
very difficult to observe.  Constraints on the fraction of PBH in this mass
range may be imposed from observations of compact stars
\cite{Capela:2012jz,Capela:2013yf,Capela:2014ita} --- neutron stars (NS) and
white dwarfs. These constraints rely on the fact that if a compact star
captures even a single PBH it gets destroyed in a short time. If an old
compact star is observed, the probability that it captures a PBH in its
lifetime must be much smaller than 1.  Thus, a mere observation of an old
neutron star in an environment with a high DM density and low velocity
dispersion of DM, where the capture rate is the highest, may imply constraints
on the fraction of PBH in the total amount of dark matter.

There are two ways to capture a PBH: during the formation of a main sequence
star, with its subsequent evolution into a compact star
\cite{Capela:2012jz,Capela:2014ita}, or through direct capture by an already
existing neutron star \cite{Capela:2013yf}. This is the second case that is
relevant for the present paper.

In order to be captured by a compact star, a PBH has to loose its kinetic
energy and become gravitationally bound to the star. Due to a microscopic size
of the PBH and large mass, the energy loss in a single passage through the
star is very small, so that only PBHs with very small asymptotic velocity can
be captured. The energy loss, therefore, becomes one of the key parameters
that determines the resulting constraints, as the amount of captured DM is
directly proportional to it.

Two different approaches have been used in the literature to calculate
the energy loss by a PBH passing through a NS. One is based on the
dynamical friction mechanism \cite{Capela:2013yf}. In this approach
the star is approximated as a collection of free particles which
interact individually with the gravitational field of a passing PBH
and  absorb part of its momentum causing the PBH to slow down. The
motivation for this approximation is that the speed with which the PBH
passed through a NS after being accelerated in the gravitational field
of the latter is at least a few times larger than the sound speed in
the NS, so that, by causality, the collective properties of the matter
do not have time to manifest themselves. The energy loss in this
approach has been calculated in Ref.~\cite{Capela:2013yf} and is,
parametrically,
\begin{equation}
E_{\rm df} \sim {G m^2\over R}\ln \Lambda,
\label{eq:Edf}
\end{equation}
where $G$ is Newton's constant, $R$ is the star radius, 
$m$ the PBH mass and the subscript
stands for dynamical friction. The factor $\ln \Lambda$ is the Coulomb
logarithm \cite{Chandrasekhar} whose presence is due to the long-range
character of the Newtonian potential. This factor is estimated as $\ln
\Lambda \sim \ln (R/Gm)$ for an ordinary star and is somewhat reduced
in the case of the neutron star because of the Fermi-degeneracy of the
nuclear matter \cite{Capela:2013yf}.  The constraints that are
obtained with this energy loss are 1-2 orders of magnitude weaker than
those resulting from the PBH capture at the stage of the star
formation \cite{Capela:2014ita}.

An alternative approach to the energy loss 
has been used in Ref.~\cite{Pani:2014rca},
where the NS matter has been treated as a medium rather than a
collection of individual particles. A passing PBH excites oscillations
of the medium by its gravitational field and thus loses the energy.
In Ref.~\cite{Ostriker} it has been shown that as far as the
excitation of the sound waves is concerned, the two approaches give
identical answer in the supersonic case, in accord with the above
causality arguments.

However, it was argued in Ref.~\cite{Pani:2014rca} that in the case of a PBH
crossing a NS, the dominant energy loss comes not from the normal sound waves,
but from the surface or gravity waves excited on the surface of the NS
core. When the energy is represented as a sum over partial waves with harmonic
number $l$, it was found that this sum diverges at large $l$, giving the
energy loss of the form \cite{Pani:2014rca}
\begin{equation}
E_{\rm sw}  \sim {G m^2 \over R} \, \sum_1^{l_{max}} {1\over l^n},
\label{eq:Esw}
\end{equation}
where $n$ depends on the matter equation of state (according to
\cite{Pani:2014rca}, $n\simeq 0.5$ for the NS core and $n = 0$ for an
incompressible fluid). When summed up to some large $l_{\rm max}$,
Eq.(\ref{eq:Esw}) gives a much larger energy loss than the dynamical friction,
Eq.(\ref{eq:Edf}), the dominant contribution coming from the shortest
wavelengths, {\em i.e.} from the vicinity of the PBH impact point. This result
is in disagreement with the causality arguments because the surface waves
propagate in any case not faster than the sound, and thus the approximation
of the dynamical friction approach should be valid. 

The aim of the present paper is to resolve this apparent 
paradox (see also \cite{Capela:2014qea}). Keeping in
mind that the enhancement factor in Eq.(\ref{eq:Esw}) comes from the vicinity
of the PBH and is maximum for stiff fluids, we developed a simple analytically
solvable model where the claims of Ref.\cite{Pani:2014rca} can be easily
tested. Namely, we consider a semi-infinite incompressible fluid in the
uniform external gravitational filed normal to the fluid surface. This model
possesses surface waves fully analogous to the waves, say, on the surface of
a lake. A PBH passing through the fluid surface excites an outgoing circular
wave that propagates from the impact point. We calculated analytically the
fluid perturbation and the amount of energy it carries to infinity which, by
energy conservation, equals the energy loss by the PBH.

We found that this energy is convergent even in the idealized case of a
point-like PBH, and parametrically coincides with Eq.(\ref{eq:Edf}), up to the
factor $\ln \Lambda$, after the appropriate identification of
parameters. Moreover, when represented in momentum space --- that is, in the
form analogous to Eq.(\ref{eq:Esw}) --- the contribution of high momenta is
cut off. The cutoff occurs precisely at the value of the momentum beyond which
the surface waves propagate slower than the PBH speed (note that, as will be
discussed below, the higher is the momentum, the slower is the propagation
speed of a surface wave). Thus, as expected, we recover the causality
arguments in the case of the surface waves as well. We also identified 
problems in the calculations of Ref.~\cite{Pani:2014rca} which we believe have
lead to an erroneous answer.

The rest of this paper is organized as follows. In section \ref{sec:planar} we
set up the basic equations and consider the gravitational interaction of a
point-like mass (the black hole) with a semi-infinite incompressible fluid in
a uniform gravitational field (a flat neutron star). From the explicit
expression of the surface deformation as a function of time, we compute the
energy transfer from the black hole to the fluid and show that it is of order
of Eq.(\ref{eq:Esw}). In section \ref{sec:sphere} we briefly consider the case
of a sphere of incompressible fluid and recover, for large momenta, the
behavior corresponding to the planar limit. In the same section we compare in
more details our result with those of \cite{Pani:2014rca} and point to
calculational issues that could explain the differences between our results. We
then draw our conclusions.

\section{Energy loss for a flat star}
\label{sec:planar}

We begin by setting up the basic equations, which may be found in
standard reference books (see, for instance,
\cite{Landau:fluids}). Since the 
fluid is incompressible  $\rho = \const$, and so is the pressure at the
surface, $p_0 = \const$. We set $p_0=0$ in what follows. The
equation of continuity applied to an incompressible fluid gives $\div
\vec v = 0$. Since we focus on gravitational effects, which derive
from a gradient, we consider that the flow is irrotational. Then the
velocity field of the fluid takes the from a gradient, $\vec v = \grad
\varphi$, with $\varphi$ satisfying
\begin{equation}
\label{eq:div}
\Delta \varphi = 0.
\end{equation}
The Euler equation in the vicinity of the boundary reduces to
\begin{equation}
\label{eq:bernouilli}
\partial_t \varphi + {1\over 2} (\grad \varphi)^2 = -{p/\rho} - g z,
\end{equation}
where $g$ is the acceleration due to gravity at the boundary. The
unperturbed surface, which is flat and of infinite extension,
corresponds to $z=0$. So we write $z = \eta(x,y,t)$ for the
deformation of the surface. To make progress we begin by neglecting the
non-linear terms. This amounts to assuming that $\eta$ is always
smaller than the characteristic wavelength of a given deformation
$\lambda$, $\vert \eta \vert \ll \lambda$. At the surface we then 
have 
\[
\left(\partial_t \varphi\right)_{z =\eta} + g \eta = 0.
\] 
To eliminate $\eta$ from this equation, we use the fact that $v_z =
\partial_z \phi \approx \partial_t \eta$, which is valid for small
deformations, and get
\begin{equation}
\label{eq:euler}
\left(\partial^2_t \varphi + g \partial_z \varphi \right)_{z=\eta} = 0.
\end{equation}
Eqs. (\ref{eq:euler}) together with Eq.(\ref{eq:div}) dictate the
evolution of small surface deformations in presence of gravity, or
gravity waves. Notice that they do not involve the density of the
fluid, which is a manifestation of the equivalence principle.

For the problem at hand, we need the eigenfunctions of
Eq.~(\ref{eq:div}) with cylindrical symmetry. To simplify the
subsequent expressions, we consider a fluid with infinite depth. We
have checked that considering finite depth does not change our
conclusions. In this limit, the velocity potential takes a simple form
in terms of the Bessel function $J_0$,
\begin{equation}
\label{eq:freesol}
\varphi_k(\vec x,t) \propto e^{-i\omega_k t} e^{kz} J_0(kr),
\end{equation}
where $z<0$. For further reference, we also present the expression for the
velocity itself,
\begin{equation}
\label{eq:velocity}
\vec v_k(\vec x,t) \propto 
e^{- i \omega_k t} e^{k z} \left (J_0(kr) 
\vec 1_z - J_1(kr) \vec 1_r\right)
\end{equation}
\[
\sim 
e^{- i \omega_k t} \vec s_k(\vec x),
\]
where $\vec s_k(\vec x)$ are the eigenmodes of the fluid displacement
vector field, see Sect.~\ref{sec:sphere}.

From this and Eq.(\ref{eq:euler}) we have
\begin{equation}
\label{eq:disrel}
\omega_k^2 = g k,
\end{equation}
which is the familiar dispersion relation of surface waves in deep water ({\em
  i.e.} $\lambda\ll h$, with $h$ the depth of the fluid layer). The group
velocity is
\begin{equation}
\label{eq:vg} 
v_g = {\partial \omega_k\over\partial k} = {1\over 2}\sqrt{g\over k}
\end{equation}
which implies that short wavelength modes travel more slowly. 

We may now consider the effects of the PBH, which we will treat as a
perturbation with gravitational potential
\[
\Phi(t,r,z) = - {G m\over \sqrt{r^2 + (z + \bar v t)^2}}
\]
where $m$ is the mass of the PBH and $\bar v$ is its velocity near the
surface. We have approximated the gravitational field of the PBH by
its Newtonian expression valid at distances $r\gg 2 G m$; the
justification will be given later. For simplicity, we also neglected the PBH
acceleration due to the gravitational field and thus consider $\bar v$
to be constant. 

With the gravitational potential included, Eq.(\ref{eq:euler}) becomes
\begin{equation}
\partial^2_t \varphi + g \partial_z \varphi = - \partial_t \Phi
\label{eq:inhomogen}
\end{equation}
at $z=0$ and we search for a solution for the velocity potential $\varphi$ 
that is zero at $t\to -\infty$. By representing $\Phi$ as a sum of
the Bessel functions in a manner similar to Eq.~(\ref{eq:freesol}), 
Eq.~(\ref{eq:inhomogen}) can be diagonalized and solved. The
result reads
\begin{eqnarray}
\label{eq:solphi2}
\varphi(t,r,z) &=& {G m \bar v\over g} \int_0^\infty {dk e^{kz} J_0(kr) \over
  1 + k\bar v^2/g} \Bigl[ -\epsilon(t) e^{-k \bar v |t|}\nonumber\\ && +
  \theta(t) 2 \cos(\omega_k t) \Bigr] , 
\end{eqnarray}
where $\epsilon(t)$ and $\theta(t)$ are the sign and Heaviside functions,
respectively. One easily checks that this function and its time derivative
are continuous at $t=0$, and that Eq.~(\ref{eq:inhomogen}) is indeed
satisfied. 

Using $\partial_z \varphi = \partial_t \eta$, we get that the
deformation of the surface is given by
\begin{eqnarray}
\label{eq:sol1}
\eta(t,r) &=& {G m\over g} \int_0^\infty \!\! dk \,{1\over 1 + k\bar v^2/g} \\
&&\times\left[ e^{ - k \bar v \vert t\vert} 
+ 2 \theta(t)\bar v \sqrt{k\over g} \sin(\omega_k t)\right] J_0(kr).\nonumber
\end{eqnarray}
For $t<0$, the deformation has the form of a bump, with a height that is
increasing as the BH comes closer to the surface.\footnote{The profile of the
  deformation is diverging at the origin at $t=0$. The divergence is mild,
  being only logarithmic, and has no ---or little--- incidence on the energy
  loss. The emergence of this divergence of is related to the point-like
  approximation of the gravitational potential of the PBH, and could be
  regulated --at the cost of substantial complication-- by taking into account the Schwarschild radius.}.  For $t>0$,
while the bump fades away, the perturbation evolves into an outgoing
wavetrain, as described by the second (oscillating) term in
Eq.~(\ref{eq:sol1}). Because the group velocity (\ref{eq:vg}) is decreasing
for large $k$, at late times this surface wave has a characteristic shape ---
typical of water surface disturbances \cite{Whitham} --- with a front ahead
and an oscillatory pattern with shorter wavelengths behind

Using Eq.(\ref{eq:sol1}), we may readily calculate the energy transfer.  To
this end we compute the energy $E$ carried away at late times by the
wavetrain, that is we consider only the contribution at $t\gg 0$ of the second
term in Eq.(\ref{eq:sol1}). This energy is given by the sum of the kinetic and
potential energy of the disturbance,
\[
E = \int {1\over 2} \rho v_{\rm wt}^2 d^3x 
+ {1\over 2} \rho g \int \eta^2_{\rm wt}  d^2x,
\]
where the subscript 'wt' stands for the part of the solution describing
the outgoing wave. 
In the second term that corresponds to the potential energy, we have
performed the integration over $z$ and subtracted the contribution of the
unperturbed fluid. From Eqs. (\ref{eq:solphi2}) and (\ref{eq:sol1}) we get
\begin{equation}
E = 4 \pi \rho {G^2 m^2 \bar v^2\over g^2} \int_0^\infty \! dk\; 
{1\over \left(1 + k \bar v^2/g\right)^2} 
\label{eq:energy-wt}
\end{equation}
\[
= 4 \pi \rho {G^2 m^2 \over g} .
\]
In Appendix~\ref{sec:energy-loss-as} we recalculate the energy
deposited into the fluid perturbations at arbitrary time, using a
different method and retaining all the terms in the solution
(\ref{eq:sol1}). We then show that the resulting expression reproduces
Eq.~(\ref{eq:energy-wt}) at $t\to\infty$.

The expression of Eq.(\ref{eq:energy-wt}) is our main result. First and
foremost we notice that Eq.(\ref{eq:energy-wt}) is UV finite, as
we advertised in the Introduction. Although this equation refers to the flat
case, the disagreement with Eq.~(\ref{eq:Esw}) is already evident from the
large-$k$ behavior: should Eq.~(\ref{eq:Esw}) be recovered, the integral over
$k$ in Eq.(\ref{eq:energy-wt}) would have to linearly diverge, which is not
the case. 

Second, only sufficiently low-$k$ modes contribute substantially to the energy
loss. The integral in Eq.(\ref{eq:energy-wt}) is cut at $k\lesssim g/\bar
v^2$. Making use of Eq.~(\ref{eq:vg}), this condition translates into
\[
v_g\gtrsim \bar v.
\]
We see that only the waves that propagate faster than the PBH are efficiently 
excited, in agreement with the causality arguments. 

Finally, by identifying the acceleration $g$ with the acceleration at the
surface of the neutron star and the velocity $\bar v$ with the velocity of the
BH falling onto it, 
\begin{equation}
g\sim {GM \over R^2}, \qquad \bar v = \sqrt{GM \over R},
\label{eq:accel}
\end{equation}
where $R$ is the radius of the star and $M$ its mass, one can cast
Eq.~(\ref{eq:energy-wt}) in the form
\begin{equation}
\label{eq:Enatural}
E \sim {Gm^2\over R}, 
\end{equation}
which is parametrically the same, apart from the logarithmic factor, as the
energy loss Eq.~(\ref{eq:Edf}) calculated in the dynamical friction
approach. A more detailed comparison with the spherical case is presented in
the next Section.

To conclude this section, let us check the validity of the
approximations used to obtain Eq.~(\ref{eq:energy-wt}). It follows
from Eq.~(\ref{eq:energy-wt}) that the dominant contribution to the
energy loss comes from the excitation of the waves with the
wavelengths 
\begin{equation}
\lambda\sim 1/k\sim {\bar v}^2/g.
\label{eq:lambda}
\end{equation}
The Newtonian
approximation for the gravitational potential of the PBH is,
therefore, valid when this wavelengths is much larger than the horizon
size of the PBH, $ {\bar v}^2/g \gg R_s=2Gm$, which is satisfied for
sufficiently small $g$ and/or large $\bar v$.  In the realistic case
of a neutron star we have from Eqs.~(\ref{eq:accel}) that ${\bar
  v}^2/g\sim R\gg R_s$, so in practice the Newtonian approximation is
satisfied for PBH lighter than about solar mass. 

Consider now the variation of the PBH velocity. At the distance of
order $\lambda$, the fractional change of the PBH velocity due to
gravitational acceleration is $\delta \bar v = g \lambda/{\bar
  v}$. From Eq.~(\ref{eq:lambda}) we conclude that $\delta \bar v \sim
\bar v$, so the constant velocity approximation is only marginally
satisfied --- enough to establish the absence of divergency and make
an order-of-magnitude estimate of the energy loss, but not enough for
a precise answer.

\section{Spherical Star and comparison of results}
\label{sec:sphere}

Armed with our understanding of the planar limit, we consider now the
case of a sphere of incompressible fluid of radius $R$. In this we
will follow the approach of \cite{Pani:2014rca} and consider the fluid
displacement vector field $\vec s(\vec x,t)$, which reduces to the
surface deformation $\eta$ for $\vec x$ at the surface. As the issue
at hand is the large-momenta behavior of the energy deposition, we
will study the limit of the solution for the sphere in the large $l$
limit, where $l$ is the multipole index.  Our motivation is
twofold. First we want to provide and independent check of the planar
solutions that we have derived. This will imply right away that the
large-$l$ modes give the same contribution to the
energy loss as in the planar case, as one should expect. In
particular, the energy loss is UV finite (see the appendix for the
expression of the energy loss for all $t$, Eqs.(\ref{eq:deltaEtm}) and
(\ref{eq:deltaEtp})). Second, we want to trace the origin of the
difference between our conclusions and those of
\cite{Pani:2014rca}. For this we compare our solution with that of 
Ref.~\cite{Pani:2014rca} in the large-$l$ limit. 

As mentioned in \cite{Pani:2014rca}, the eigenproblem for a sphere of
an incompressible gravitating fluid has been solved by Kelvin.
Expressing the eigenfunctions in terms of spherical harmonics, for the
sphere of radius $R$ and mass $M$, the fluid
displacement eigenmodes  $s_{l}(\vec x)$ are given by
\begin{equation}
\label{eq:disp_sphere}
\vec s_{l}(\vec x) = \sqrt{4 \pi l \over 3 M} 
\left({\tilde r \over R}\right)^{l-1} \left(Y_{l0}(\theta) 
\vec 1_{\tilde r} +{1\over l} \partial_\theta Y_{l0}(\theta) 
\vec 1_\theta\right)
\end{equation}
where $\tilde r$ is the radial coordinate $0 < \tilde r < R$. We will
assume that the PBH trajectory is passing through the
center of the sphere, so the only relevant modes correspond to
$m=0$. Their normalization is as in \cite{Pani:2014rca}. The 
frequencies of the eigenmodes are given by
\begin{equation}
\label{eq:omega_sphere}
\omega_l^2 =  {2 l(l-1)\over 2 l +1} {G M\over R^3}
\end{equation}
The frequency depends only on the multipole index $l\geq 2$ because
there are no excited radial modes for an incompressible sphere. Using
$g = G M/R^2$, the dispersion relation becomes Eq.(\ref{eq:disrel})
for large $l \approx k R$. Similarly, using
\[
Y_{l,0}(\theta, \varphi ) = \sqrt{2 l + 1\over 4 \pi} \, 
P_l(\cos\theta) \approx \sqrt{k R\over 2 \pi} J_0(k r)
\]
valid for large $l \approx k R$, we obtain the expression of the
eigenmodes in the planar limit
\begin{equation}
\label{eq:disp_planar}
\vec s_l(x) \sim {k \over\sqrt{ 2 \pi \rho R}} e^{kz}\left(J_0(kr) 
\vec 1_z - J_1(kr) \vec 1_r\right).
\end{equation}
Up to an arbitrary normalization factor, this agrees with
Eq.(\ref{eq:velocity}).

Now we claim that the solution for the fluid displacement for all time in the
presence of the PBH is given by
\begin{equation}
\label{eq:sol_gen}
\vec s(\vec x,t) = \sum_{l=2}^\infty \vec s_l(\vec x)\int_{-\infty}^t 
dt^\prime  A_l(t^\prime) {\sin\omega_l(t-t^\prime)\over \omega_l}
\end{equation}
where 
\begin{equation}
\label{eq:Alref}
A_l(t) = \int_{\rm star} dV \, \vec f\cdot \vec s_l(x),
\end{equation}
$\vec f(x)$ being the gravitational force of the PBH acting on a fluid
element in the point $\vec x$.  This is pretty clear from the fact
that, in the sector with given $l$,
\[
G^+_l(t-t^\prime) = \theta(t-t^\prime) 
{\sin\omega_l(t-t^\prime)\over \omega_l}
\]
is the retarded Green's function for the problem at hand, with
$\omega_l$ given by Eq.(\ref{eq:omega_sphere}). This solution,
Eq.(\ref{eq:sol_gen}), differs from the expression used in
\cite{Pani:2014rca}, a point to which we will come back later on.

First we consider the large $l \approx k R$ limit of the
solutions. Using the force decomposition in spherical modes in the
large-$l$ limit 
\begin{eqnarray}
\vec f &=& - G m \rho\int_0^\infty dk k\, e^{- k \vert z+\bar vt\vert} \nonumber\\
&& \times \left(\epsilon(z+\bar v t) J_0(kr) \vec 1_z + J_1(kr) \vec 1_r\right)
\end{eqnarray}
and Eq.~(\ref{eq:disp_sphere}) we have
\begin{equation}  
\label{eq:Al}
A_l(t) \sim \sqrt{2 \pi \rho\over R}  G m\, e^{- k \bar v \vert t\vert }
\end{equation} 
from which  it is  straightforward to get
\begin{equation}
\label{eq:stp1}
\vec s(\vec x,t< 0) 
= G m\int \!dk\,{e^{k(z+\bar v t)}\over g 
+ k \bar v^2}\left(J_0(kr ) \vec 1_z - J_1(kr) \vec 1_r\right)
\end{equation}
and
\begin{eqnarray}
\label{eq:stp}
\vec s(\vec x,t> 0) &=& G m \int \!dk\,\left({e^{-k\bar v
    t}\over g 
+ k \bar v^2}- {2 \bar v  k \over g + k \bar v^2}
{\sin\omega_kt\over \omega_k}\right) \nonumber\\
&&\times e^{kz} \left(J_0(kr ) \vec 1_z - J_1(kr) \vec 1_r\right)
\end{eqnarray}
This solution gives precisely the surface deformation given by
Eq.(\ref{eq:sol1}). Alternatively one may check that it corresponds to the
velocity field resulting from Eq.(\ref{eq:solphi2}), with $\vec v = \grad
\varphi \equiv {d \vec s/dt}$.

Our first conclusion is thus that the two approaches ---the direct resolution
of the planar problem and the large $l$ limit of the spherical problem--- lead
to precisely the same solutions, as they should.  This not only provides an
independent check of our approach, but also shows that the large $l$ behavior
on the sphere is mundane, the same as in the planar limit. From this we also
conclude that the energy loss is finite. Indeed, the only divergence we have
encountered in the planar limit is for small momenta, but this is regulated by
the radius of the sphere $R$. Unfortunately, full calculations beyond the
large-$l$ limit, in particular that of the energy loss, are not as
straightforward as in the planar case. Nevertheless, in the light of the
discussion of the previous section, the final result should be similar to that
of Eq.(\ref{eq:Enatural}).

We do not reach the conclusions of \cite{Pani:2014rca}, so what is the source
of the disagreement?  We believe that our result is sound, essentially it
matches the intuition that there should be little quantitative difference
between energy loss through tidal deformation and dynamical friction. Indeed,
as we have argued above, the underlying mechanism is the same, and both
approaches should match for supersonic propagation, as in
\cite{Ostriker}. Regarding the calculations, the comparison of the solutions
in the large-$l$ limit reveals several 
calculational issues that may explain the difference between our results and
those of \cite{Pani:2014rca}. 

For the sake of simplicity we focus on Eq.(2.8) in \cite{Pani:2014rca}, which
together with their equations (2.2) and (2.3) should in principle give the
fluid displacement for $t<0$. In the large $l$, corresponding to the planar
limit, we get from their result (reinstating $G$ and setting $\bar v$ as the
velocity near the surface)
\begin{eqnarray}
\label{eq:s-PandL}
\vec s(\vec x,t<0)\vert_{P\&L} &= &{G m} \int \! dk\, \sqrt{R k\over 2 \pi}\, 
{\bar v^2 k\over g}\, {1 \over g + k \bar v^2} \\
&& \times e^{k (z + \bar v t )} \left (J_0(kr) 
\vec 1_z - J_1(kr) \vec 1_r\right).\nonumber
\end{eqnarray} 
This solution, obtained from the intermediate expressions in
Ref.~\cite{Pani:2014rca}, differs from our solution,
Eq.(\ref{eq:stp1}). In particular, it has a much worse large-$k$
behavior. 

The first issue is the presence of the factor of $ \sqrt{ k
  R/2 \pi} $ which we believe should be absent, since the planar limit
(if taken appropriately) should contain no explicit factors of
$R$. Re-doing the steps in \cite{Pani:2014rca} that lead to their
Eqs.(2.8) and (2.9) one indeed realizes that a factor of 
$\sqrt{4 \pi / (2 l +1)}\sim \sqrt{2 \pi/ k R}$ is missing. We do not know whether this a mere misprint, or if this missing factor has propagated in the calculations of the energy loss. 

Another problem is the presence of the factor $\bar v^2 k/ g$ compared to our
Eq.(\ref{eq:stp1}), which leads to the an extra divergence in $k$ of the
energy loss obtained in \cite{Pani:2014rca}.  We believe that the resolution
to this specific issue is as follows. First it is clear Eqs.(\ref{eq:sol_gen})
and (\ref{eq:Alref}) are undoubtedly correct.  However, the solution in
\cite{Pani:2014rca} is obtained using a distinct starting point, given by
their Eqs.~(2.2) and (2.3). To see the relation, we integrate by part
Eq.(\ref{eq:sol_gen}) to get
\begin{equation}
\label{eq:byparts}
\vec s(\vec x,t) =\mbox{\rm Re}\sum_{l=2}^\infty {\vec s_l(\vec  x)\over 
\omega_l^2}\int_{-\infty}^t dt^\prime \partial_{t^\prime} A_l(t^\prime) 
\left(1 -{e^{i\omega_l(t-t^\prime)}}\right).
\end{equation}
The first term in the brackets is the boundary term.  Going from this
expression to that of Eqs. (2.2) and (2.3) in \cite{Pani:2014rca} involves two
further steps \cite{Dahlen:1998a}. First, one must neglect the boundary term. This may be fine if
the source (here the PBH gravitational field) is far enough from the star, but
leads to a wrong expression for the fluid deformation at any finite
time. Second, one must assume that the derivative over time and the
integration over space may be exchanged. These are subtle effects, which 
are difficult to spot in the full-fledged spherical system, so it is again
useful to consider the planar limit.

Let us deal first with the boundary term. Using Eqs.~(\ref{eq:byparts}) and
(\ref{eq:Al}), we get
\begin{eqnarray}
\vec s(\vec x,t<0) &=& G m \, \mbox{\rm Re}  
\int dk {k \over \omega_k^2} \left(1 - {\bar v k \over k 
\bar v - i \omega_k}\right)  \\
&\times&  e^{k z + k \bar v t}\left(J_0(kr) 
\vec 1_z - J_1(kr) \vec 1_r\right),
\nonumber
\end{eqnarray}
where the $1$ in the first brackets comes from the boundary term in
Eq.~(\ref{eq:byparts}).  If taken alone, the term $\propto \bar v k$ under the
brackets gives the result of \cite{Pani:2014rca} (without the factor of
$\sqrt{k R/2 \pi}$ discussed in the previous paragraph). Including the
boundary term (the $1$ under the bracket) gives instead our result,
Eq.(\ref{eq:stp1}), which has a much milder large $k$ behaviour. We believe
that this settles this specific problem.

We have considered so far the case $t<0$, which is most transparent, and found
an extra factor of $k^{3/2}$ in the solution of Ref.~\cite{Pani:2014rca} as
compared to our result. This factor is also present in their solution at
$t>0$. However there is one extra issue, which is that in calculating
$\partial_t A_l(t)$  it is assumed in \cite{Pani:2014rca} (and in
\cite{Dahlen:1998a}) that the time derivative and the volume integration
commute,
\[
\partial_t \int_{\rm star} dV \, \vec f\cdot s_l(\vec x) 
\stackrel{?}{=} \int_{\rm star} dV \,  \partial_t \vec f\cdot s_l(\vec x).
\]
To see that this is not the case for the problem at hand it is again useful to
take the large-$l$ limit. From Eq.(\ref{eq:Al}) we have
\begin{equation}
\label{eq:us}
\partial_t \int_{\rm star} dV \, \vec f\cdot s_l(\vec x)  
\propto \epsilon(t) e^{-k \bar v \vert t \vert}
\end{equation}
while calculating the derivative and the integral in a different order --- the
one of Ref.~\cite{Pani:2014rca} --- gives instead
\begin{equation}
\label{eq:pani}
\int_{\rm star} dV \, \partial_t \vec f\cdot s_l(\vec x)  
\propto  e^{-k \bar v \vert t \vert}.
\end{equation}
The appearance of the sign function $\epsilon(t)$ in the correct expression
(\ref{eq:us}) is crucial to get our result for the fluid displacement,
Eq.(\ref{eq:stp}). One may check that failing to take into account this change
of sign leads to an expression for the fluid displacement for $t>0$ that
diverges linearly for large $k$.  Together with the factor of $\sqrt{k R/2
  \pi}$ discussed above, one would get a result consistent with the large $l$
limit behaviour of the $t>0$ solution in \cite{Pani:2014rca}, which we think
is incorrect and do not reproduce here to avoid further cluttering of
equations.

To summarize, the fluid displacement in the spherical case should have a
large-$l$ limit given by Eqs.~(\ref{eq:stp1})--(\ref{eq:stp}) which is
consistent with the planar case. However, the solution of
Ref.~\cite{Pani:2014rca} has a different (more singular) behavior in $k$,
which may explain the appearance of divergency in the resulting energy loss at
large $l$.

\section{Conclusions}
Motivated by the problem of energy loss by a PBH passing through a
neutron star and the controversy existing in the literature, we have
considered a simple model where the crucial questions concerning this
phenomenon can be addressed analytically. The model consists of an
infinite incompressible fluid in a uniform external gravitational
field normal to the fluid surface, so that the fluid is in 
equilibrium. When perturbed, this system possesses surface waves
reminiscent of those on the surface of a lake. 

We have calculated the excitation of these waves by a gravitational
field of a moving mass (say, a PBH) crossing normally the fluid
surface. From the explicit solution for the fluid perturbations we have
calculated the energy transferred to the fluid and, therefore, lost by
the passing mass. We have found that: 
\begin{itemize} 
\item The transferred energy Eq.~(\ref{eq:energy-wt}) is convergent at
  high momenta even in the limit of a point-like mass. Our result is
  thus in disagreement with that of Ref.\cite{Pani:2014rca},
  Eq.~(\ref{eq:Esw}).
\item The contributions into the energy transfer of individual waves
  with a given momentum $k$ is cut at high values of momentum. The
  cutoff corresponds to the suppression of modes that propagate (much)
  slower than the velocity of the passing mass. This supports the
  causality arguments in the case when surface waves are present.
\item Upon the appropriate identification of the parameters, our
  resulting energy transfer (\ref{eq:energy-wt}) is parametrically the
  same as the dynamical friction result, apart from the logarithmic
  factor. The absence of the Coulomb logarithm in the flat
  infinite case suggests that this factor would also be absent in the
  realistic spherical case. 
\item The generalization of our model to the case of a spherical ball
  of an incompressible gravitating fluid is difficult to completely
  solve analytically. Nevertheless, the detailed comparison of the
  flat and spherical cases allows one to identify a problem in the
  calculation of Ref.\cite{Pani:2014rca}.
\end{itemize}

Even though a real neutron star is neither flat nor incompressible,
our calculation can still shed some light on the relative importance
of the surface wave contribution into the energy loss. First, the
absence of the Coulomb logarithm suggest that this contribution is
subdominant compared to the dynamical friction one, at least by the
logarithmic factor. Second and more important, we see from
Eq.~(\ref{eq:energy-wt}) that contributions of slow (high-$k$) modes
into the transferred energy are suppressed, because these modes are
not excited efficiently. We think this is a general phenomenon not
related to such peculiar features of our model as incompressibility
of the fluid. The suppression factor is easy to estimate from
Eq.~(\ref{eq:energy-wt}). The modes propagating with velocities $v\ll
\bar v$ contribute to the transferred  energy (\ref{eq:energy-wt}) the
fraction of order $(v/\bar v)^2$. 

The PBH falling onto a neutron star attains a velocity which is about
$0.5 c$. The speed of the surface waves at the boundary of the NS core
depends on their wavelength, but in any case cannot exceed the sound
speed, which is at least an order of magnitude smaller than the speed
of the PBH. We thus expect that the energy transfer into the
excitation of the surface waves in the case of a realistic NS is
suppressed by a factor of $(v/\bar v)^2\lesssim 10^{-2}$ as compared
to the dynamical friction result. A full calculation in the
case of a spherical ball of a compressible fluid is needed to verify
this expectation, which goes beyond the scope of this paper.

\appendix

\section{Alternative derivation of the energy loss}
\label{sec:energy-loss-as}
It is
instructive to calculate 
the energy loss by the BH as a function of time in the planar limit.  For this,
following \cite{Press:1977a}, we calculate the work done by the BH on the
fluid elements,
\begin{equation}
{d E\over dt} =  \int d^3 x \vec v \cdot \vec f_{BH} 
\end{equation}
where $\vec v = \grad \varphi$ is the velocity field of the fluid and $\vec
f_{BH} = - \grad \Phi$. 
Integrating over $t$, we get the energy transferred $\Delta E$ as a function
of time:
\begin{equation}
\label{eq:deltaEtm}
\Delta E(t<0) = 2 \pi G^2 m^2 \rho\int_{k_{co}}^\infty\! {dk\over 2 k (g + k \bar v^2)} e^{2 k \bar v t}
\label{eq:energy-work1}
\end{equation}
\begin{equation}
\label{eq:deltaEtp}
\Delta E(t>0) = 2 \pi G^2 m^2 \bar v \rho \int_{k_{co}}^\infty\!  {dk\over g + k \bar v^2}\left\{{e^{-2 k
    \bar v t}\over 2 k \bar v} + \right.
\label{eq:energy-work2}
\end{equation}
\[
\left. + {2 \over k^2 \bar v^2 + \omega_k^2}\left [ k \bar v + e^{- k \bar vt}(\omega_k\sin(\omega_kt)- k \bar v\cos(\omega_kt)\right]\right\}
\]
 It is straightforward to see that at
positive times the term that does not contain the exponential factor exactly
reproduces Eq.~(\ref{eq:energy-wt}). The other terms are logarithmically
divergent at small $k$, which reflects the long-range character of the
Newtonian potential. When cut at $k_{co} \sim 1/R$, $R$ being the star radius, they
decay with time starting from the value parametrically given by
Eq.~(\ref{eq:energy-wt}). Thus, at $t\to \infty$ the work performed by the BH
on the fluid tends to the energy stored in the outgoing waves.

\acknowledgments 
We thank Fabio Capela for helpful discussions during the early stage of this project. We also thank Paolo Pani for discussions. 
 Our work is
supported by the IISN, an ULB-ARC grant and by the Belgian Federal
Science Policy through the Interuniversity Attraction Pole P7/37
``Fundamental Interactions''. M.T. also acknowledges support and hospitality from the LPT at Universit\'e d'Orsay, Paris-Sud.

\end{document}